\documentclass[twocolumn,showpacs,preprintnumbers,amsmath,amssymb]{revtex4}

\usepackage{graphicx}
\usepackage{dcolumn}
\usepackage{bm}

\begin{document}

\preprint{APS/123-QED}

\title{A Millennium Scale Sunspot Number Reconstruction: Evidence For an Unusually Active Sun Since the 1940's}
\author{Ilya G. Usoskin}%
 \email{Ilya.Usoskin@oulu.fi}
\affiliation{Sodankyl\"a Geophysical Observatory (Oulu unit), FIN-90014 University of Oulu, Finland}%
\author{Sami K. Solanki}
\email{Solanki@linmpi.mpg.de}
 \affiliation{Max-Planck Institut f\"ur Aeronomie, Katlenburg-Lindau, Germany}
\author{Manfred Sch\"ussler}
\affiliation{Max-Planck Institut f\"ur Aeronomie, Katlenburg-Lindau, Germany}%
\author{Kalevi Mursula and Katja Alanko}%
\affiliation{Department of Physical Sciences, University of Oulu, Finland}%

\date{\today}

\begin{abstract}
The extension of the sunspot number series backward in time is of
considerable interest for dynamo theory, solar, stellar, and climate research.
We have used records of the $^{10}$Be concentration in polar ice to reconstruct
the average sunspot activity level for the period between the year 850 to the present.
Our method uses physical models for processes connecting the $^{10}$Be concentration with the sunspot
number.
The reconstruction shows reliably that the period
of high solar activity during the last 60 years is unique throughout the
past 1150 years. This nearly triples the time interval for which such a
statement could be made previously.
\end{abstract}

\maketitle

\section{Introduction}
The sunspot number (SN) series represents the longest running direct
record of solar activity, with reliable observations starting in 1610,
soon after the invention of the telescope.  The behaviour of solar
activity in the past, before the era of direct measurements, is of
importance for a variety of reasons.  For example, it allows an improved
knowledge of the statistical behaviour of the solar dynamo process which
generates the cyclically varying solar magnetic field.  It also should
help to produce superior estimates of the fraction of time the Sun
spends in states of very low activity, the so-called Great Minima, such
as the Maunder Minimum in the second half of the 17$^{\mbox{th}}$
century. This is of particular interest when comparing the behaviour of
the Sun with that of other Sun-like stars \cite{bali90}.  The level of
solar activity also affects the Sun's radiative output \cite{will91},
which in turn may influence the Earth's climate \cite{sola98}.  However,
any such influence takes place on time scales longer than the solar
cycle, so that a statistically significant comparison with paleoclimatic
records requires a long time series of solar activity data.

We are specifically interested in the past evolution of sunspot
activity. Sunspots lie at the heart of solar active regions and trace
the emergence of large-scale magnetic flux, which is responsible for the
various phenomena of solar activity. Consequently, sunspots are a good
tracer for solar magnetic activity, particularly so during times of
medium to high activity.

The sunspot number record shows intriguing contrasts between the
extremes reached during the Maunder Minimum when practically no
sunspots were seen on the face of the Sun \cite{eddy76,ribe93}, and in
the period since the 1940s when SN reached the average value of about 75.
Extensions to earlier times have been attempted in the past by
extrapolating this record, based on mathematical modelling using
statistical properties of the observed SN record \cite{nago97,rigo01} or
adjusting them to fragmentary data on naked-eye sunspot and auroral
observations \cite{shov55}.  Such extrapolations suffer from rapidly
increasing uncertainty for earlier times. Alternatively, the SN prior to
1610 has been estimated from archival proxies, such as the concentration
of cosmogenic $^{14}$C isotope in tree rings or $^{10}$Be isotope in ice cores
drilled in Greenland and Antarctica \cite{beer90,damo91,beer00}. For
want of a physical relationship, a simple linear regression between the
SN and the isotope concentration has generally been assumed.

Recently, detailed physical models have been developed for each
individual link in the chain connecting the SN with the cosmogenic
isotopes. This includes a physical model relating the heliospheric
magnetic flux (the Sun's open magnetic flux) to the SN
\cite{sola00,sola02}, a model for the transport and modulation of
galactic cosmic rays within the heliosphere \cite{usos02a}, and a model
describing the $^{10}$Be isotope production in the terrestrial
atmosphere \cite{webb03,masa99}.  We have combined these models, such
that the output of one model becomes the input for the next step.  It
has thus become possible to model the complete sequence of processes and
to calculate the expected $^{10}$Be concentration from 1610 onwards on
the basis of the SN record \cite{usos02b,usos03}.  The inversion of this
chain has been successfully demonstrated as well \cite{usos03}.
For artificial, noise-free $^{10}$Be data the yearly SN could be reconstructed
with an error of $\pm10$ compared with the typical solar maximum SN of over 100
 during recent cycles.  For real $^{10}$Be data, the noise makes the
reconstruction of yearly sunspot numbers impractical, but robust
reconstructions of 11-year (solar cycle) averages of this quantity are
shown to be possible.

\begin{figure}[t]
\centering
\resizebox{\hsize}{!}{\includegraphics{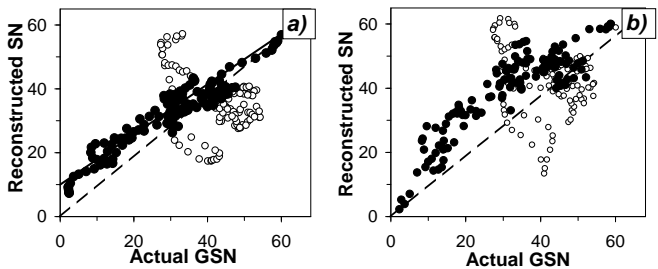}}
\caption{Scatter plots of {annual} sunspot numbers (SN) reconstructed
 from the {11-year smoothed} Greenland Dye-3 $^{10}$Be data
 \cite{beer90} vs. the {annual}, 11-year smoothed group sunspot number
 for the period 1700-1940. {\em Left:} results from our physical
 reconstruction method. We find a nearly linear relation (solid dots)
 with a small amount of scatter, except for four excursions (open dots)
 during {specific} periods of time. These stronger deviations are
 probably caused by climatic effects.
 {\em Right:} results of a fit based upon linear
 regression between group sunspot number and $^{10}$Be production
 rate. Our physical model yields a much closer relation between the
 actual and the reconstructed SN than the purely statistical approach.
}
\end{figure}

Follow the recent approach \cite{usos03} we present here a
reconstruction of the SN since the year 850, based upon the measured
$^{10}$Be concentrations in ice cores at the Dye-3 site in Greenland
 (annual data for 1424--1985) \cite{beer90} and at the South Pole
 (roughly 8-year sampled data for 850--1900) \cite{bard97}.

\section{Sunspot number reconstruction}

The reconstruction comprises five steps, which are described in detail by
Usoskin et al. \cite{usos02b,usos03}:
\begin{equation}
^{10}{\rm Be}\stackrel{(1)}{\longrightarrow} CR\stackrel{(2)}{\longrightarrow} \Phi \stackrel{(3)}{\longrightarrow}
F_o\stackrel{(4)}{\longrightarrow}S\stackrel{(5)}{\longrightarrow}\,SN
\label{Eq:model1}
\end{equation}
First the flux of cosmic rays (CR) impinging on the Earth's atmosphere is derived
 from the measured $^{10}$Be concentration.
This yields the modulation strength $\Phi$ of cosmic rays through a model of
 heliospheric transport of cosmic rays \cite{usos02a}, which in turn is used to
 determine the Sun's open magnetic flux $F_o$.
The model \cite{sola00,sola02} of Solanki et al. is employed to obtain the source term
 $S$ for the open magnetic flux and, finally, the sunspot number SN.
We use a realistic $^{10}$Be yield function \cite{webb03} and
include also heavier species of cosmic rays, in particular
$\alpha$-particles, whose contribution to the $^{10}$Be
production is about 30\%.  The advantage of our physical
reconstruction is that it takes into account the non-linear
nature of the relation between $^{10}$Be concentration and SN and
thus returns more reliable values of SN once the parameters of
the physical models are fixed on the basis of actual measurements.
For the period of time from 1700 to 1940 when both SN and
 $^{10}$Be data sets are
 reliable, Fig.~1 illustrates the improvement
 achieved with our reconstruction method (left panel) compared to a
 linear regression between the SN and the $^{10}$Be production rate
 (right panel).
The distinct excursions over limited time
 intervals (open circles: 1722--1743, 1761--1789, 1828--1845,
 1882--1894) are probably due to local climatic effects in
 Greenland, as indicated by comparison with the $^{14}$C data (see Fig.~2),
 which are much less affected by local climatic variability.
Most of the time, however, the reconstructed SN is closely related to the actual SN
 observations \cite{note1}, with the full dots showing a
considerably smaller scatter {(with a linear correlation coefficient of
$r=0.96$)} than the corresponding result using a linear regression
{($r=0.81$)}.
While our reconstruction reproduces the SN at
 intermediate to high levels of solar activity relatively well,
 it tends to overestimate the SN during periods of low activity.
This {partly} reflects the
existence of a residual level of magnetic activity which modulates the
 cosmic ray flux even in times of almost vanishing sunspots
 \cite{beer98,usos01}.
Such activity could, e.g., be due to the emergence
of small, spotless ephemeral active regions \cite{harvey92,sola02}.
The SN based on our physical reconstruction can therefore be considered as
upper limits during periods of low SN.
The assumption that $^{10}$Be is deposited locally \cite{usos03} may also
 contribute to the offset at small SN seen in Fig.~1.
Mixing in the terrestrial atmosphere would cause the $^{10}$Be concentration
 to be affected by geomagnetic variations.
Since the Earth's magnetic field was stronger during the
 Maunder minimum than it is today, neglecting this change could lead
 to an overestimate of the reconstructed SN.
We have used the (nearly linear) relationship between the actual and the
reconstructed values on the left panel of Fig.~1 to apply a simple
correction to the reconstruction, which mainly affects the low SN
values.

\begin{figure*}[t]
\centering
\resizebox{\hsize}{!}{\includegraphics{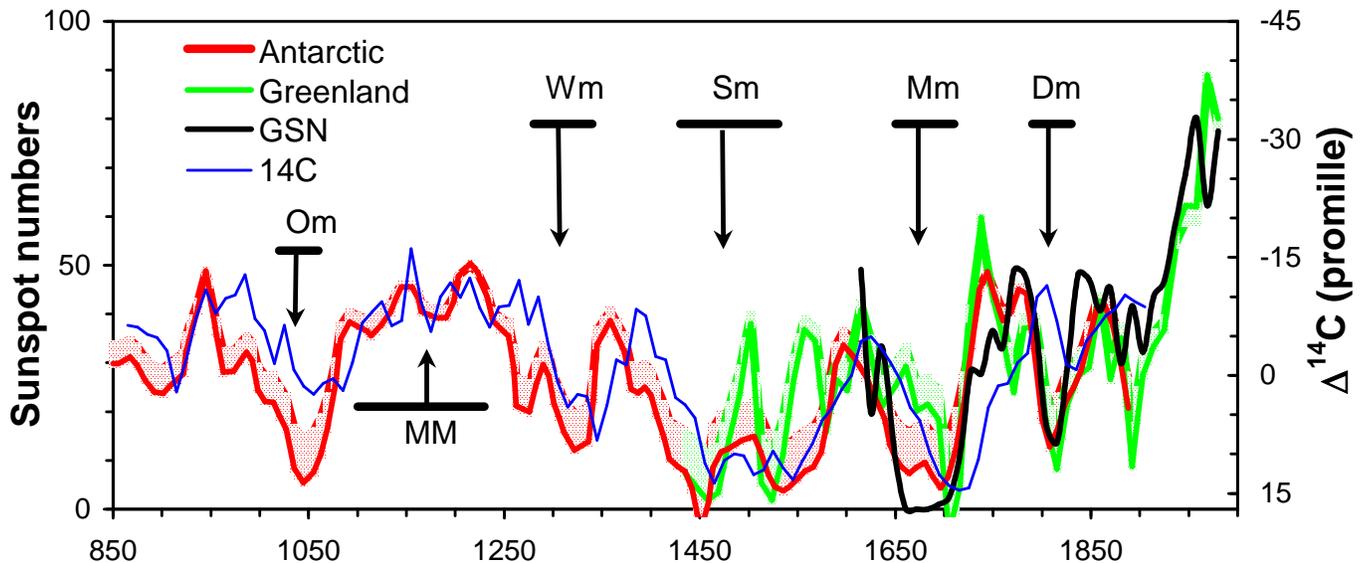}}
\caption{Time series of the sunspot number as reconstructed from
$^{10}$Be concentrations in ice cores from Antarctica (red) and
Greenland (green). The corresponding profiles are bounded by the actual
reconstruction results (upper envelope to shaded areas) and by the reconstructed
values corrected at low values of the SN (solid curves) by taking into
 account the residual level of solar activity in the limit of vanishing
 SN (see Fig.~1).
The thick black curve shows the
 observed group sunspot number since 1610 and the thin blue curve gives the
(scaled) $^{14}$C concentration in tree rings, corrected for the
variation of the geomagnetic field \cite{stui80}.
The horizontal bars
with attached arrows indicate the times of Great Minima and Maxima
\cite{stui89}: Dalton minimum (Dm), Maunder minimum (Mm), Sp\"orer
minimum (Sm), Wolf minimum (Wm), Oort minimum (Om), and Medieval Maximum
(MM).  The temporal lag of $^{14}$C with respect to the sunspot number
is due to the long attenuation time for $^{14}$C \cite{bard97}.}
\end{figure*}

\section{Discussion}

Fig.~2 shows the (1-2-1 averaged) SN reconstructed from the 8-year
 sampled Antarctic $^{10}$Be record for the years 850--1900
 \cite{bard97} and from the Greenland Dye-3 record for the period
 1424--1985 \cite{beer90}.
Also given is the similarly averaged group sunspot
number \cite{hoyt98} based on observations after 1610 and the (scaled)
$^{14}$C concentration in tree rings, corrected for the change of the
geomagnetic field \cite{stui80,stui89}.
For easier comparison, the latter curve has been scaled to match
 the mean and the range of the reconstructed SN.
The reconstructed SN profiles {depicted by the (red
and green) coloured areas} are bounded {from above} by the actual
reconstruction results and from below by the low-SN corrected
results described above.

The reconstructed SN series confirms the various Great Minima and also
the Medieval Maximum (roughly between the years 1100 and 1250)
identified in previous, statistical studies of the $^{10}$Be and
$^{14}$C records \cite{stui89,bard97}. The two reconstructed and the
measured SN series generally are in good agreement after the end of the
Maunder minimum around 1700.
The differences between the results from
 the Antarctica and the Greenland $^{10}$Be records are greater
 in 1450--1700, during the so-called `Little Ice Age' \cite{eddy76},
 and they can possibly be ascribed to local climatic effects in $^{10}$Be deposition.
This interpretation is supported by the good correlation between the reconstruction
 from the Antarctica data and the $^{14}$C record during this period, once the phase
 shift of about 20 years due to the long attenuation time for $^{14}$C \cite{bard97}
 has been taken into account.

The most striking feature of the complete SN profile is the uniqueness
of the steep rise of sunspot activity during the first half of the 20th
century.  Never during the eleven centuries prior to that was the Sun
nearly as active.  While the average value of the reconstructed SN
between 850 and 1900 is about 30, it reaches values of 60 since 1900 and
76 since 1944.  For the observed group SN series since 1610 these values
are 25, 61, and 75, respectively.  The largest 100-year average of the
reconstructed SN prior to 1900 is 44, which occurs in 1140--1240, i.e.,
during the Medieval Maximum, but even this is significantly less than
the level reached in the last century.
The Medieval Maximum is remarkable, however, in the length of time that
 the Sun has consistently remained at the average SN level of about 40--50.
Only during the recent
period of high activity since about 1830, i.e., after the Dalton
minimum, has the SN remained consistently above 30 for a similar length
of time. We conclude that the high level of solar activity since the
1940s is unique since the year 850. This can be considered a robust
conclusion since we have shown that our reconstruction is particularly
reliable in phases of high and intermediate sunspot activity, { while
during periods of low activity the SN may be overestimated.}

The good overall agreement of the reconstructed SN with the $^{14}$C data
 further supports the reliability of our reconstruction: the cross-correlation
 coefficient (taking into account the overall 20-year delay in
 $^{14}$C concentration) is $0.83\pm 0.07$. Since the globally mixed
$^{14}$C is not affected by the vagaries of the local climate, the
good correspondence between the $^{14}$C curve and the SN
reconstructed from the Antarctic $^{10}$Be data indicates that
long-term climatic variability does not strongly affect our
results. This conclusion is reinforced by the good agreement between
the measured $^{14}$C concentrations and the corresponding values
derived from the $^{10}$Be data on the basis of a $^{14}$C
redistribution model \cite{bard97}.
The fact that the reconstruction based on the Greenland Dye-3 core
shows stronger deviations from $^{14}$C during the Little Ice Age
suggests that the Greenland $^{10}$Be record is more strongly affected
by local climate fluctuations \cite{bard97}.

It is known that the geomagnetic field has decreased by about 30\%
during the last 1000 years (see, e.g., \cite{baum98}).  The stronger geomagnetic
field in earlier times has led to a more effective shielding of cosmic
rays and may, depending on the amount of atmospheric mixing of $^{10}$Be
before precipitation, have caused a reduced $^{10}$Be production.  Our
calculations use the present geomagnetic field and neglect the possible
effect of the changing geomagnetic field. This is probably well
justified at least for the Antarctica record \cite{usos03}, { as
indicated by the good correspondence with the $^{14}$C record, which has
been corrected for changes of the geomagnetic field \cite{stui80}.}  In
any case, without such correction our reconstruction model ascribes any
effect of a stronger geomagnetic field to a higher SN in the
past. Consequently, our reconstructed SN values in the pre-telescopic
era are to be considered as upper bounds, emphasizing even more the
exceptional nature of the high solar activity during the last 60 years.

Although our SN reconstruction still covers a rather limited length of
time (but nonetheless about 3 times longer than the telescopic sunspot
record), the unusually high number of sunspots during the past century
suggests that we currently may be seeing a state of the solar dynamo
that is uncharacteristic of the Sun at middle age. Also, the higher
activity level implies more coronal mass ejections and more solar
energetic particles hitting the Earth.  Thus we expect that the late
20th century has been particularly rich in phenomena like geomagnetic
storms and aurorae.  The flux of energetic galactic cosmic rays (in the
neutron monitor energy range above {several} GeV) reaching the Earth is
presently about 10\% lower than it was around 1900 \cite{usos02b}.  The
suppression of lower-energy cosmic rays (about 2 GeV), which are mainly
responsible for the production of cosmogenic isotopes, is even stronger,
reaching up to 40\%.

The current high level of solar activity may also have an impact on
the terrestrial climate. We note a general similarity between our
long-term SN reconstruction and different reconstructions of
temperature \cite{mann99,jone01}: (1) both SN and temperature
show a slow decreasing trend just prior to 1900, followed by a
steep rise that is unprecedented during the last millenium; (2) Great
Minima in the SN data are accompanied by cool periods while the
generally higher levels of solar activity between about 1100 and 1300
correspond to a relatively higher temperature (the Medieval Warm
Period) \cite{brad01}.  To clarify whether this similarity reflects a
real physical connection requires a more detailed study of the
various proposed mechanisms for a solar influence on climate
\cite{issi00}.

We thank the anonymous referee for useful comments on improving
 this paper.


\begin{thebibliography}{}
\bibitem{note1} {During the period 1722-1743, the reconstructed SN is
closer to the Wolf SN series than to the group SN.}
\end{thebibliography}
\end{document}